\documentclass[twocolumn,aps,prc,tightenlines,floats,floatfix]{revtex4}

\def\sfrac#1#2{{\textstyle \frac{#1}{#2}}}

\def\bea{\begin{eqnarray}}
\def\eea{\end{eqnarray}}

\def\be{\begin{equation}}
\def\ee{\end{equation}}
\def\ba{\begin{eqnarray}}
\def\ea{\end{eqnarray}}

\usepackage{graphics}
\usepackage{graphicx}
\usepackage{epsf} 
\usepackage{amsmath}
\usepackage{amssymb}

\begin{document} 

\phantom{0}
\vspace{-0.2in}
\hspace{5.5in}
\parbox{1.5in}{ \leftline{JLAB-THY-09-940}}

\vspace{-1in}

\title
{\bf Valence quark contribution for the $\gamma N \to \Delta$ 
quadrupole transition extracted from lattice QCD}
\author{G. Ramalho$^{1,2}$ and M.T. Pe\~na$^{2,3}$ 
\vspace{-0.1in}  }

\affiliation{
$^1$Thomas Jefferson National Accelerator Facility, Newport News, 
VA 23606, USA \vspace{-0.15in}}
\affiliation{
$^2$Centro de F{\'\i}sica Te\'orica de Part{\'\i}culas, 
Ave.\ Rovisco Pais, 1049-001 Lisboa, Portugal \vspace{-0.15in}}
\affiliation{
$^3$Department of Physics, Instituto Superior T\'ecnico, 
Ave.\ Rovisco Pais, 1049-001 Lisboa, Portugal}

\vspace{0.2in}
\date{\today}

\phantom{0}

\begin{abstract} 
Starting with a covariant spectator quark model developed
for the nucleon ($N$) and the $\Delta$ in the physical pion mass
region, we extend the predictions
of the reaction $\gamma N \to \Delta$
to the lattice QCD regime.
The quark model includes
S and D waves in the quark-diquark wavefunctions. Within this framework
it is the D-wave part in the $\Delta$ wavefunction that
generates nonzero valence contributions for the
quadrupole form factors of the transition.
Those contributions are however insufficient to explain
the physical data, since the pion cloud
contributions dominate. To separate the two effects
we apply the model to the
lattice regime in a region where
the pion cloud effects are
negligible, and adjust the D-state
parameters directly to the lattice data. 
This process allows us to obtain a better determination
of the D-state contributions.
Finally, by adding a simple parametrization
of the pion cloud we establish the
connection between the experimental data and the lattice data.
\end{abstract}

\vspace*{0.9in}  
\maketitle

\section{Introduction}
\label{secI}

In recent years, the structure 
of the baryon resonances has been an important 
topic of both experimental and theoretical investigation.
Of particular interest is the $\Delta$ resonance, 
the first excited state of the nucleon.
New $\Delta$ photo-production for large four-momentum transfer have 
been extracted in several laboratories,
as Jlab, MAMI, LEGS and MIT-Bates \cite{CLAS,CLAS06,MAMI,LEGS,Bates}.
Simultaneously, lattice simulations 
were performed for both
the $\gamma \Delta \to \Delta$ \cite{LatticeDeltaFF,Aubin} and the
$\gamma N \to \Delta$ \cite{Leinweber93,Alexandrou,Alexandrou08}
transition form factors.     

This last transition 
can be described in terms of the 
Jones and Scadron 
multipole form factors \cite{Jones73}:
the magnetic dipole M1 ($G_M^\ast$),
the electric E2 ($G_E^\ast$), and the
Coulomb C2 ($G_C^\ast$) quadrupole form factors.
The reaction is dominated by the magnetic dipole form factor $G_M^\ast$.
Although constituent quark models (CQM) 
\cite{NDelta,NDeltaD,Becchi65,Isgur82,Giannini07,Capstick95,Riska,DeSanctis04,Braun06},
with valence quark degrees of freedom only, 
are sufficient to explain some properties 
of the $\Delta$,  
they are not sufficient to explain the
$G_M^\ast$ data at the physical 
point ($m_\pi = m_\pi^{phy}= 138$ MeV) 
\cite{NDelta,NDeltaD,Pascalutsa06b}.
The inclusion of chiral symmetry 
and/or a coupling of the quark to the 
pion field \cite{Faessler06,Lu97,BuchmannEtAl,LiRiska}
helps to overcome the limitations of the pure valence quark models.
The importance of the pion field, or "pion cloud", was also 
demonstrated using effective field theory 
\cite{Pascalutsa06,Gail06,Arndt04}
and model reaction mechanisms based on the hadronic fields,
known as dynamical models \cite{DMs,Drechsel07,Diaz07a}.
See Ref.~\cite{Pascalutsa06b} for a review 
of the $\gamma N \to \Delta$ transition.

In previous works we successfully described  the 
nucleon \cite{Nucleon,FixedAxis}, the $\gamma \Delta \to \Delta$ \cite{DeltaFF}
and the $\gamma N \to \Delta$\cite{NDelta,NDeltaD} transitions, by
considering a spectator quark model 
inspired by the Vector Meson Dominance (VMD) mechanism.
When we restrict the $\Delta$ wavefunction to S-waves,
the spectator valence quark model gives
no contributions to the quadrupole form factors $G_E^\ast$ and $G_C^\ast$
\cite{NDelta}.
Those contributions emerge only when D-states
are considered, consistent 
with other CQMs \cite{Becchi65,Isgur82}.
The D-states
improve the description of the 
experimental data, but their contributions are in general small, 
with the quadrupole form factors at the 
physical point being dominated by the 
pion cloud ~\cite{NDeltaD} (see also \cite{Giannini07,Pascalutsa06b} 
for a review).
The dominance of the pion cloud at the physical point 
prevents an accurate calibration of the valence quark 
D-state contributions and, consequently,
the relative amount of the contributions from
D-waves only and the pion cloud
is not yet exactly known \cite{NDeltaD,Pascalutsa06b,Drechsel07}.

The VMD mechanism, which we used to parametrize the 
electromagnetic interaction in terms 
of the hadronic masses, can also be used to extend the 
covariant spectator model to the lattice regime.
By introducing the dependence of the hadronic masses on the lattice 
pion mass we could describe 
the lattice data for the nucleon \cite{Gockeler05}
and for the magnetic dipole 
form factor $G_M^\ast$ \cite{Alexandrou08} of the 
$\gamma N \to \Delta$ reaction, with only S-waves
in both the nucleon and the $\Delta$ wavefunctions \cite{Lattice}.
This motivated that
in the present work we include D-states in the $\Delta$ in wavefunction, 
to show whether the
description of the lattice data is 
still possible, also for the $G_E^\ast$ and
$G_C^\ast$ form factors.

In order to include the D-states, and to overcome
the uncertainty discussed above on their effects,
we start by realizing that the valence quark effects are
expected to dominate 
for large pion masses, exactly when 
pion cloud effects are 
expected to be suppressed \cite{Detmold}.
In light of this, lattice QCD, in particular 
in quenched approximation with $m_\pi > 400$ MeV, 
where the pion cloud effects
are negligible  \cite{Detmold}, becomes the ideal laboratory 
to test the valence quark contributions, and to 
constrain the D-states. 
Importantly, the dependence of the two sub-leading
$G_E^\ast$ and
$G_C^\ast$ form factors on the
D-states allows then the extraction of
very relevant information. 

In this work we use the lattice information to
determine more precisely the valence quark effects
in the $\gamma N \to \Delta$ quadrupole form factors, and
consider the contributions of the 
valence quark structure, including orbital D-states, in the regime of the
lattice calculations.
First, we started by making a direct application of
the valence quark
model fixed in the physical region 
($m_\pi = m_\pi^{phy}$) in Ref.~\cite{NDeltaD},
to the quadrupole moments in the lattice region.
Although the model generates the correct order of 
magnitude of the lattice quadrupole data,
it fails to reproduce their $Q^2$ dependence.
Therefore, we inverted the procedure:
we began by readjusting the $\Delta$ model parametrization 
to the quenched lattice data,
imposing an overall description of 
the lattice data for several values of the pion mass. 
From the resulting
parametrization we generated directly
the valence quark contribution of the orbital
D-states to $G_{E}^\ast$ and ${G_C^\ast}$ at the 
physical point.
Finally, 
by adding a pion cloud contribution 
derived in the large-$N_c$ limit,
which was established independently from our model,
we obtain a successful description of 
the experimental data.

The paper is organized as follows:
in Sec.~\ref{formalism}
we review the formalism associated with 
the spectator quark model;
in Sec.~\ref{extension} we explain 
how to generalize the model to the 
lattice QCD regime; 
in Sec.~\ref{lattice} we present the results 
of that generalization;  
in Sec.~\ref{physical} we present 
the predictions for the quadrupole 
form factors in the physical region
and discuss the results;
finally in Sec.~\ref{conclusions}
we draw conclusions.

\section{Formalism}
\label{formalism}

We consider a quark model 
based on the covariant spectator formalism \cite{Gross,Gross06}.
In this formalism
the nucleon and the $\Delta$ 
are described as a system of  
an off-mass-shell quark and two non-interacting quarks 
forming an on-mass-shell diquark
\cite{NDelta,NDeltaD,Nucleon,Gross06}. 

In our model, the quarks are effective degrees of freedom
"dressed" by form factors, and the nucleon and $\Delta$ wavefunctions 
are not derived from a dynamical wave equation, 
but given a parametric form consistent with the intrinsic symmetries 
of these systems. By construction \cite{Nucleon,NDelta}
the wavefunctions are covariant and reproduce
their expected non-relativistic limits.
The first feature is important for the 
application of the formalism to the
kinematics of the recent data.

The nucleon 
and $\Delta$ 
wavefunction can be expressed in terms 
of a quark spin (isospin) state together with
a spin-0 (isospin-0) diquark state and a spin-1 (isospin-1) vector diquark 
state \cite{Nucleon,FixedAxis,NDelta}, multiplied by
a relative angular momentum state \cite{NDeltaD}.

The electromagnetic transition 
current between the nucleon and the $\Delta$ 
can be expressed \cite{Nucleon,FixedAxis,NDelta,NDeltaD} as
\be
J^\mu= 3 \sum_{\lambda} \int_k \bar \Psi_\Delta(P_+,k) 
j_I^\mu \Psi_N(P_-,k),
\label{eqJ1}
\ee
where $j_I^\mu$ is a generic isospin dependent 
quark current, $\Psi_\Delta$, $\Psi_N$ 
are respectively the wavefunction of the 
nucleon (momentum $P_-$) and $\Delta$ (momentum $P_+$).
The hadronic current (\ref{eqJ1}) involves the sum 
in all intermediate diquark polarizations $\lambda=0,\pm 1$, and
the invariant 
integral $\int_k \equiv \int \frac{d^3 k}{2 E_s(2\pi)^3}$ over the
diquark momentum,
where $E_s$ is the diquark on-mass-shell energy
$E_s=\sqrt{m_s^2+{\bf k}^2}$ 
($m_s$ is the diquark mass).
The factor 3 accounts for the flavor symmetry.

For the quark current we consider the general form 
\be
j_I^\mu= j_1 \gamma^\mu + j_2   \frac{i \sigma^{\mu \nu} q_\nu}{2 m_N}, 
\label{eqJI}
\ee
where $m_N$ is the nucleon mass 
and $j_1, j_2$ contain the quark form factors, which depend on the transferred
four-momentum  squared $Q^2$. They can be 
decomposed into isoscalar ($+$) and isovector ($-$) operators 
that act in the baryon isospin states according to
\be
j_i(Q^2)= \sfrac{1}{6}f_{i+}(Q^2) + 
 \sfrac{1}{2}f_{i-} (Q^2)\tau_3.
\ee
To represent the electromagnetic quark 
form factors $f_{i\pm}$ ($i=1,2$) we consider a 
parametrization inspired by VMD.
In particular, following \cite{Nucleon,NDelta,NDeltaD} we used
\ba
\hspace{-.8cm}
& &
f_{1\pm}(Q^2)=  \lambda + (1-\lambda) 
\frac{m_v^2}{m_v^2+ Q^2} +
c_{\pm}\frac{Q^2 M_h^2}{\left(M_h^2+Q^2\right)^2}
\label{eqF1}
\\
\hspace{-.8cm}
& &
f_{2\pm}(Q^2)= \kappa_\pm 
\left\{ 
d_\pm \frac{m_v^2}{m_v^2+ Q^2} +
(1-d_{\pm}) \frac{Q^2 }{M_h^2+Q^2} \right\}.
\label{eqF2}
\ea
In this parametrization $\lambda$ was adjusted to give  
the charge number density in the deep inelastic limit \cite{Nucleon},
$m_v$ represents a vector meson
($m_v=m_\rho,m_\omega$), $M_h$ is a mass 
of an effective  heavy vector meson 
simulating the 
short-range structure, 
and $\kappa_+$ ($\kappa_-$) the isoscalar (isovector) 
quark anomalous moments. 
These two last values were adjusted  
to reproduce the nucleon magnetic moment 
$\kappa_+=1.639$ and $\kappa_-=1.823$ \cite{Nucleon}.
The coefficients 
$c_\pm$ and $d_\pm$ 
were adjusted to reproduce the nucleon 
electromagnetic form-factors.
In the calculation presented here we took the
current parametrization corresponding to model II in Ref.~\cite{Nucleon}
for the nucleon elastic form factors. 
The explicit values of the VMD 
coefficients are 
$c_+=4.16$, $c_-=1.16$ and  $d_\pm=-0.686$.
For the effective heavy meson we took 
$M_h=2m_N$.
The values for $\lambda$ and the diquark mass 
are 1.21 and $m_s=0.87 \, m_N =817$ MeV, respectively.

For both the nucleon and the $\Delta$ we consider 
wavefunctions that contain the correct
spin-isospin structure, with orbital parts
modeled by scalar functions $\phi$ of the 
quark four-momentum squared $(P-k)^2$, 
expressed in terms of the ratio
\be
\chi_H = \frac{(m_H-m_s)^2 -(P-k)^2}{m_H m_s},
\label{eqChi}
\ee
where $m_H$ represents the nucleon or the $\Delta$ mass
($m_N$ and $m_\Delta$).

For the nucleon, we consider 
an S-state wavefunction \cite{Nucleon,FixedAxis,NDelta},
including a mixture of a spin-0 and isospin-0 
with spin-1 and isospin-1 diquark structure.
In particular we used  
\be
\phi_N (P,k) = 
\frac{N_0}{m_s(\beta_1+\chi_N)(\beta_2 + \chi_N)}, 
\label{eqPsiN}
\ee
where $N_0$ is a normalization constant.
The parameters $\beta_1$ and $\beta_2$ ($\beta_2 > \beta_1$)
can be interpreted as Yukawa mass parameters.
Then $\beta_2$ accounts for the short range 
physics and $\beta_1$ for the long range.
The corresponding parameters 
can be found in Refs.~\cite{Nucleon,NDelta,NDeltaD}.

As for the $\Delta$, we consider an admixture of 
S and D states, where, as explained in Ref.~\cite{NDeltaD}, the
D-state wavefunction decomposes into a spin 1/2 core (D1 state) 
and a spin 3/2 core (D3 state), both with isospin 3/2: 
\be
\Psi_\Delta=N\left[ \Psi_S + a \Psi_{D3} 
+ b \Psi_{D1} \right],
\label{eqPsiD}
\ee
where $a$ and $b$ are admixture coefficients, 
$\Psi_S$ represents the (symmetric) $\Delta$ S-state,
and the remaining two possible D states.
The normalization constant becomes $N= 1/\sqrt{1 + a^2+ b^2}$.
To represent the momentum 
probability distribution 
of the quark-diquark $\Delta$ system we 
have, as in \cite{NDeltaD},
\ba
& &
\phi_S= \frac{N_S}{m_s(\alpha_1+\chi_\Delta)^3} 
\label{eqPsiS} \\
& &
\phi_{D3} = \frac{N_{D3}}{m_s^3(\alpha_2+\chi_\Delta)^4} \\
& &
\phi_{D1} = 
\frac{N_{D1}}{m_s^3} \left\{
\frac{1}{(\alpha_3+\chi_\Delta)^4} - 
\frac{\lambda_{D1}}{(\alpha_4+\chi_\Delta)^4} 
\right\}. 
\label{eqPsiD1}
\ea 
Similar to the nucleon case, $\alpha_i$ are 
momentum range parameters, and $N_X$ normalization constants.
Since
with the inclusion of the D-states 
the S-state can be parametrized with only 
one range parameter ($\alpha_1$), as shown in Ref.~\cite{NDeltaD},  
we relabeled all the range parameters, 
with $\alpha_1$ standing for the 
average of the values of the best model of Ref.~\cite{NDeltaD}.
In expression  (\ref{eqPsiD1}), the coefficient $\lambda_{D1}$ 
was chosen to impose 
the orthogonality between the nucleon S-state 
and the $\Delta$ D1 state
(see Ref.~\cite{NDeltaD} for details).
The extra power in Eq.~(\ref{eqPsiS}) when 
compared to the corresponding equation for the nucleon Eq.~(\ref{eqPsiN}),
was introduced to take into 
account the $G_M^\ast$ falloff observed 
at low and intermediate $Q^2$ \cite{NDelta,NDeltaD,Pascalutsa06b}.
The dependence of the D-states on the ratio $\chi_\Delta$ 
was chosen to reproduce the 
behavior of the $\gamma N \to \Delta$ 
transition \cite{NDeltaD} in perturbative QCD. 

All the wavefunctions are normalized 
in order to reproduce the nucleon \cite{Nucleon,NDelta} 
and $\Delta$ \cite{NDelta,NDeltaD,DeltaFF} charge.
In particular, one has \cite{NDeltaD}:
\ba
& &
\int_k \left[\phi_N(\bar P,k) \right]^2 = \int_k 
\left[\phi_S(\bar P,k)\right]^2=1 \\
& &
\int_k \left[ \tilde k^2
\phi_{D3}(\bar P,k) \right]^2 = 
\int_k \left[ \tilde k^2
\phi_{D1}(\bar P,k) \right]^2 = 1,
\ea
where $\bar P$ represents the momentum 
in the respective baryon frame $\bar P=(m_H,0,0,0)$.
The variable $\tilde k$ defined 
as  $\tilde k= k - \frac{P \cdot k}{m_H^2}P$ 
was introduced in Ref.~\cite{NDeltaD} to 
represent the D-states.

\begin{figure}[t]
\centerline{\mbox{
\includegraphics[width=2.4in]{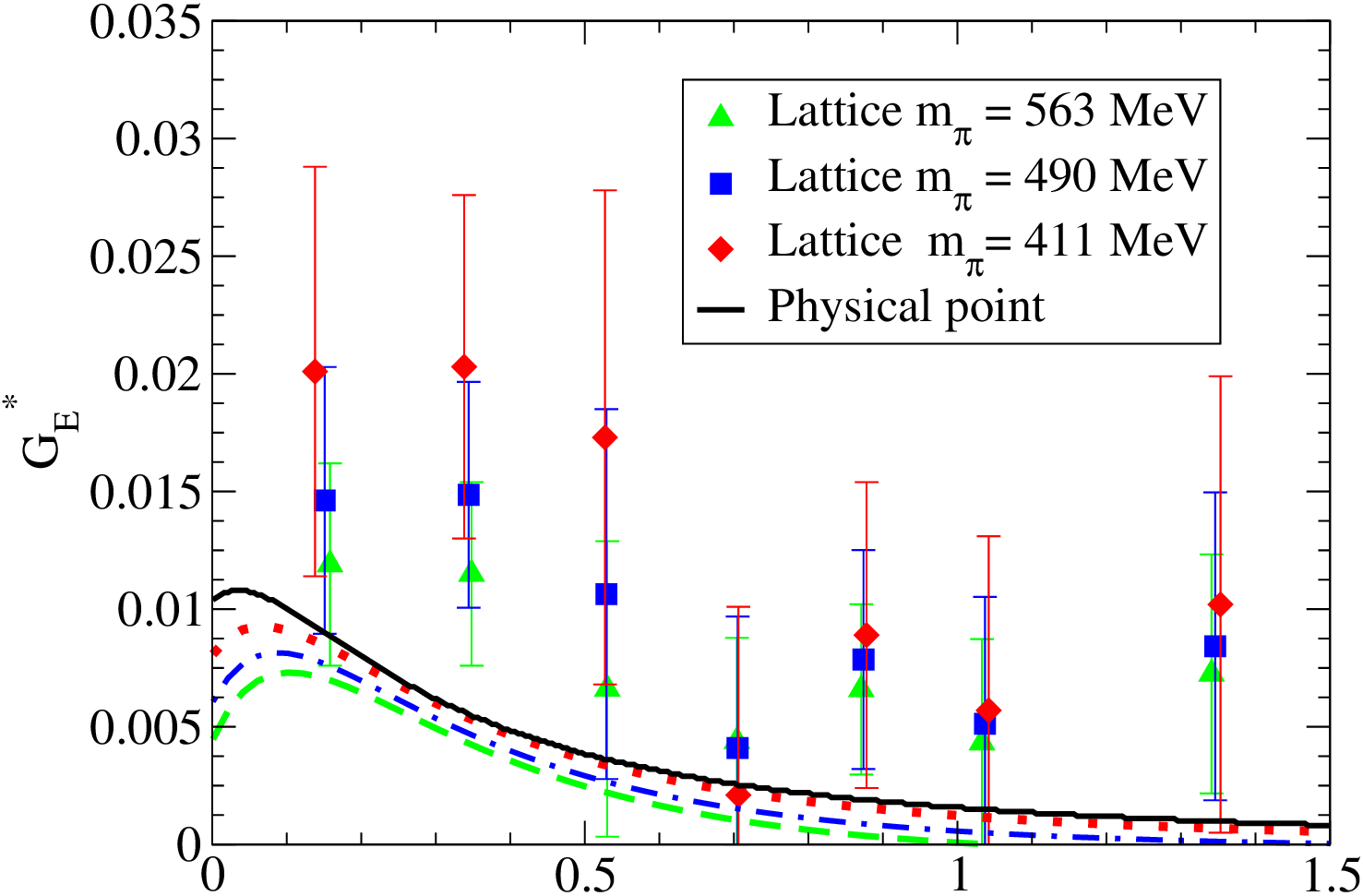}}}
\vspace{0.8cm}
\centerline{\mbox{
\includegraphics[width=2.4in]{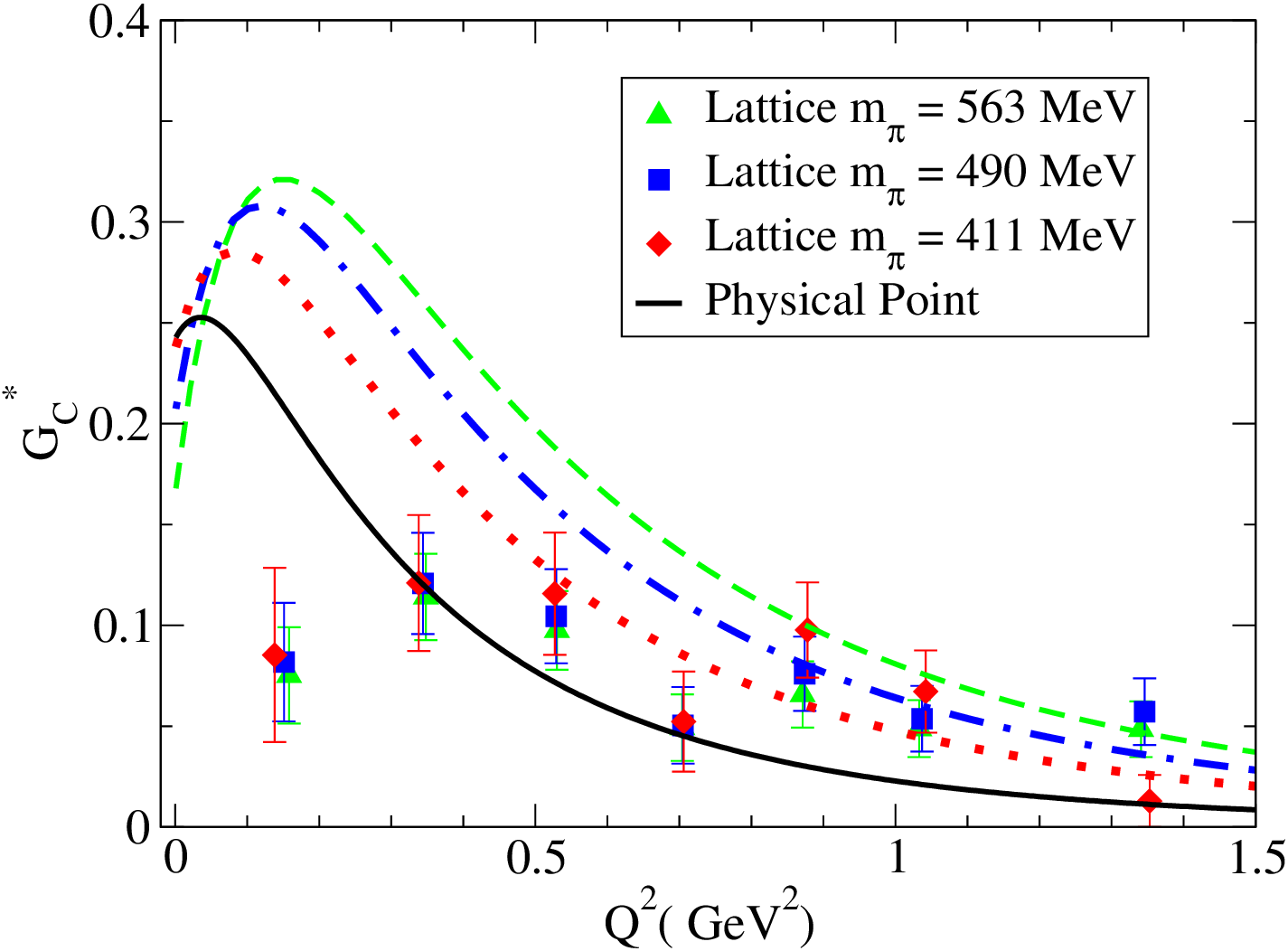}}}
\caption{Quadrupole form factors in  
quenched lattice QCD \cite{Alexandrou08}. 
The lines corresponds to the parametrization  
of the model 4 from Ref.~\cite{NDeltaD}.   
The lattice data corresponds to  
$m_\pi=$ 563 MeV (dashed line), 
490 MeV (doted-dashed line) and 411 MeV (doted line).
The solid line is the valence
quark contribution in the physical point
(same result as in Ref.~\cite{NDeltaD}).}
\label{figModelII}
\end{figure}

In summary, the nucleon (S-state) wavefunction 
contains two range parameters ($\beta_1$ and $\beta_2$),
the $\Delta$ S-state wavefunction  
a single parameter ($\alpha_1$),
and the $\Delta$ D-states contain three
($\alpha_i$, with $i=2,3,4$) parameters.
The D-state admixture is controlled additionally by the
$a$ and $b$ parameters defined
in Eq.~(\ref{eqPsiD}). 
All the parameters associated to the
quark current and the nucleon wavefunction were fixed by the
nucleon elastic data
at the physical point in a previous work \cite{Nucleon}.
They are kept unchanged here,
where we fit only five parameters to the lattice QCD data:
the three range parameters and the two admixture
coefficients for the two D-states.

\section{Extension of the spectator model 
to the lattice regime}
\label{extension}

\begin{table}[t]
\begin{center}
\begin{tabular}{c c c c c c c}
\hline
$m_\pi$   & &  $m_N$ & & $m_\Delta$ & & $m_\rho$   \\
\hline 
0.563	 & & 1.267  & &  1.470 & &  0.898 \\
0.490    & & 1.190  & &  1.425 & &  0.835 \\
0 411	 & & 1.109  & &  1.382 & &  0.848 \\
\hline
0.138    & & 0.939  & &  1.232 & &  0.776 \\ 
\hline
\end{tabular}
\end{center}
\caption{Masses in GeV, considered in lattice QCD 
\cite{Alexandrou08} and at the physical point.}
\label{tabMasses}
\end{table}

Now we consider the extension of the model presented 
in the previous section, and based  on the valence quark degrees 
of freedom, to the
lattice QCD data region.
As the pion cloud effects are 
expected to be suppressed for $m_\pi > 400$ MeV 
\cite{Detmold}, it is justified  in this region to consider
the valence contributions only. 
Therefore, we generalize the original (valence) spectator model  
by considering the implicit dependence 
of all the mass parameters on the pion mass.
Since the wavefunctions
depend on the ratio defined in Eq.~(\ref{eqChi})
the diquark mass 
scales out from the current and consequently 
from the form factors  \cite{Nucleon}.
The dependence of the model 
on the pion mass appears through
both the quark current and the baryon wavefunctions.
The quark form factors (\ref{eqF1})-(\ref{eqF2}) 
in the current operator (\ref{eqJI}) 
are written 
in terms of a VMD parametrization, and therefore
depend on a vector meson mass
$m_v$ and an effective heavy meson mass $M_h=2 m_N$. 
The extension to lattice is done by
considering the nucleon and $\rho$ masses from the lattice
calculation as an implicit function of $m_\pi$.

The  nucleon and $\Delta$ wavefunctions 
(\ref{eqPsiN}) and (\ref{eqPsiS})-(\ref{eqPsiD1}) 
are represented in terms 
of the (adimensional) momentum range  parameters
($\beta_i$ and $\alpha_i$) and the  
kinematic ratio $\chi_H$ of Eq.~(\ref{eqChi}).
Because the mass dependence of the wavefunction 
enters in this ratio,
we expect only a weak dependence on the range 
parameters ($\alpha_i$ and $\beta_i$) 
near the physical region. 
In fact, this weak sensitivity
of the range parameters to the pion mass was already verified
in the work of Ref.~\cite{Lattice}
for light pions.
For this reason we use
the same range parameters both in the lattice data region  
and at the physical point,
neglecting any pion mass dependence 
of the range parameters in the considered 
region ($m_\pi < 600$ MeV).
For heavier pions, the interaction becomes almost 
pointlike, and at least the short-range coefficients
should 
be corrected, or an explicit dependence 
on the pion mass introduced.

The procedure presented here corresponds to 
the case of Ref.~\cite{Lattice} with $M_\chi= + \infty$
($M_\chi$ is the constituent quark mass in the chiral limit).
A more detailed treatment is possible 
with finite values of  $M_\chi$, but 
for the quadrupole lattice data the corrections are small
when compared to the statistical errorbands 
and the differences between the 
datasets associated with different values for the pion masses.

To start with we took the best valence quark
model presented in Ref.~\cite{NDeltaD}, 
fixed for the physical data case 
($m_\pi =m_\pi^{phy}$).
The analytical expressions for $G_M^\ast$, 
$G_E^\ast$ and $G_C^\ast$ are presented in Ref.~\cite{NDeltaD}.
No readjustment of the parameters
of the quark current and of
the nucleon and $\Delta$ wavefunctions was made,
except for the nucleon, $\Delta$
and $\rho$ masses, which
are explicit functions of $m_\pi$, as in the lattice
calculations.
The mass parameters are presented in Table
\ref{tabMasses}.

Figure \ref{figModelII} shows the results 
of model 4 of Ref.~\cite{NDeltaD} 
extended to the quenched lattice QCD calculations of
Ref.~\cite{Alexandrou08}. We only show $G_E^\ast$ and $G_C^\ast$, since
$G_M^\ast$ does not change much from \cite{Lattice}.
The conclusion is that we cannot reproduce the lattice data 
accurately, but the predictions of the model 
have the right order of magnitude. 
The poor description of $G_E^\ast$ and $G_C^\ast$, obtained from taking
the physical model to the lattice domain, contrasts with what happens
for the dominant $G_M^\ast$ form factor obtained in Ref.~\cite{Lattice}. 
It shows that
the lattice data is more sensitive to the  D-wave components
of the $\Delta$ wavefunction, crucial
for the quadrupole form factors, than to the
S-wave components, that dominate $G_M^\ast$.
This makes
the lattice data for the quadrupole form factors extremely interesting,
as a potential source of information
on the D-wave effects in the hadronic structure, and thus indirectly,
on the magnitude of the pion cloud effects. 
In Ref.~\cite{NDeltaD}, when the model was applied
to the physical point, the valence quark 
contributions were not explicitly 
separated from the pion cloud contributions.
For that reason the estimate of the
valence contribution could not be made very precise, 
as we confirm in this extension to the 
lattice regime.

\begin{figure*}[t]
\centerline{
\mbox{
\includegraphics[width=2.3in]{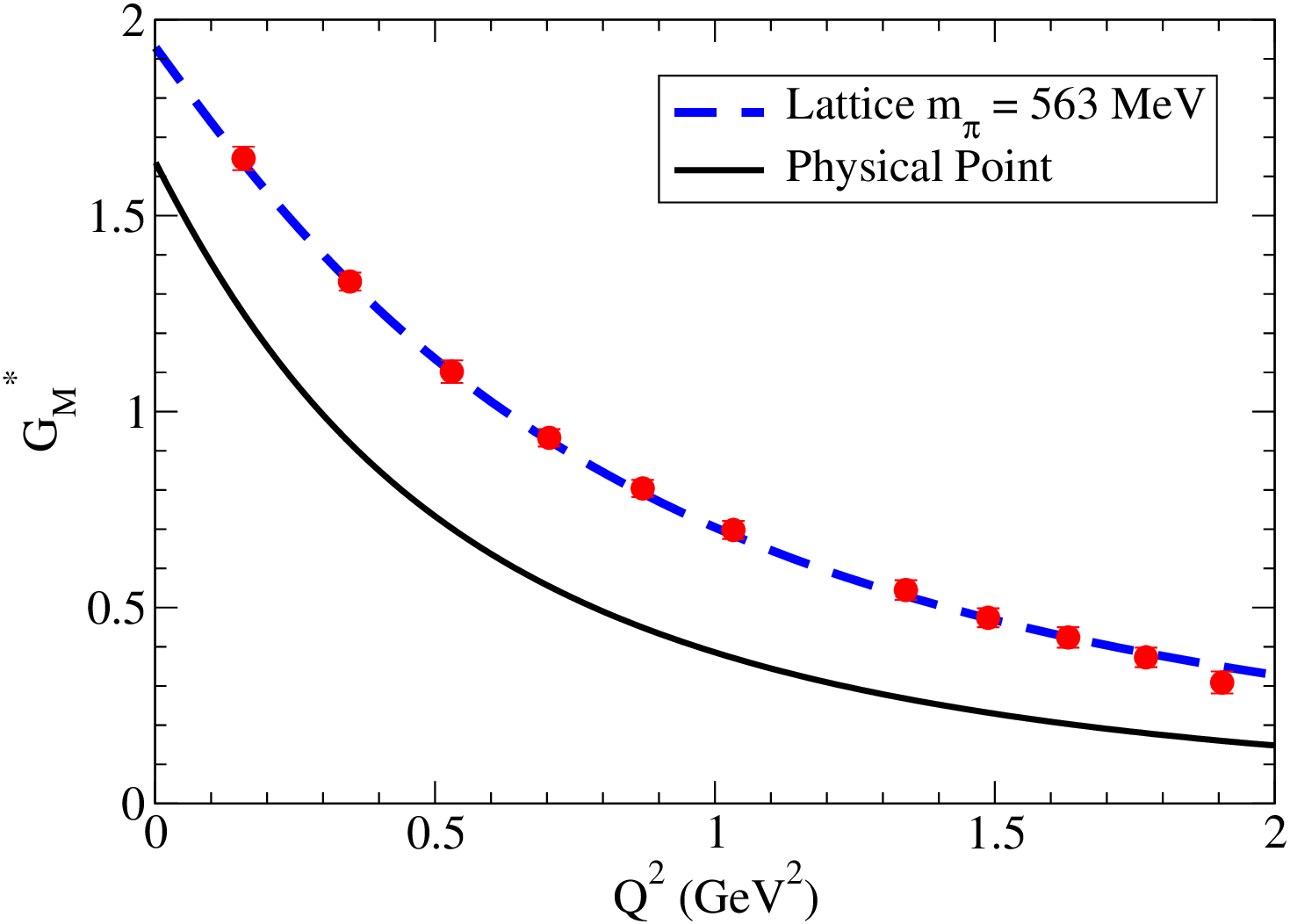}
\includegraphics[width=2.3in]{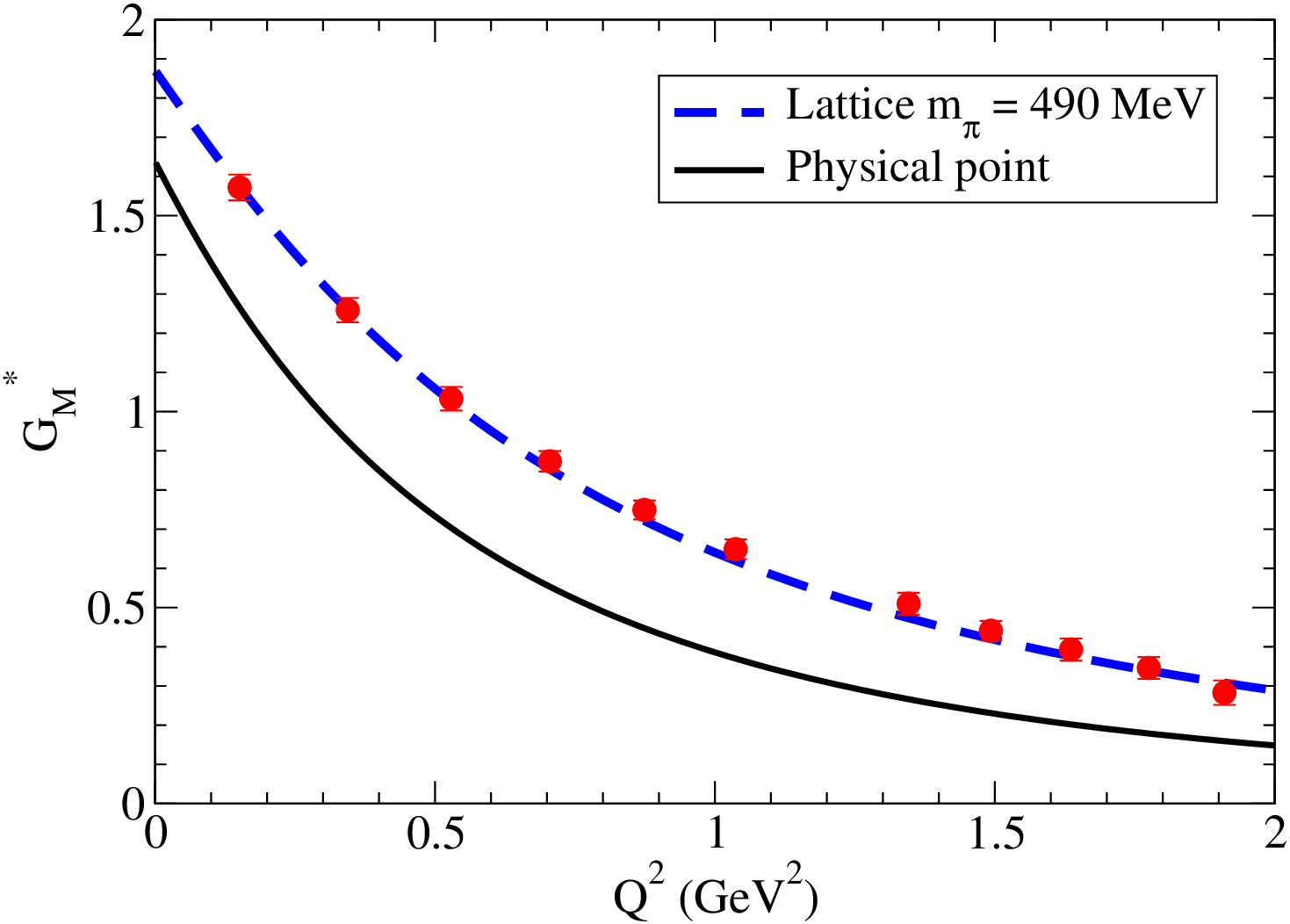}
\includegraphics[width=2.3in]{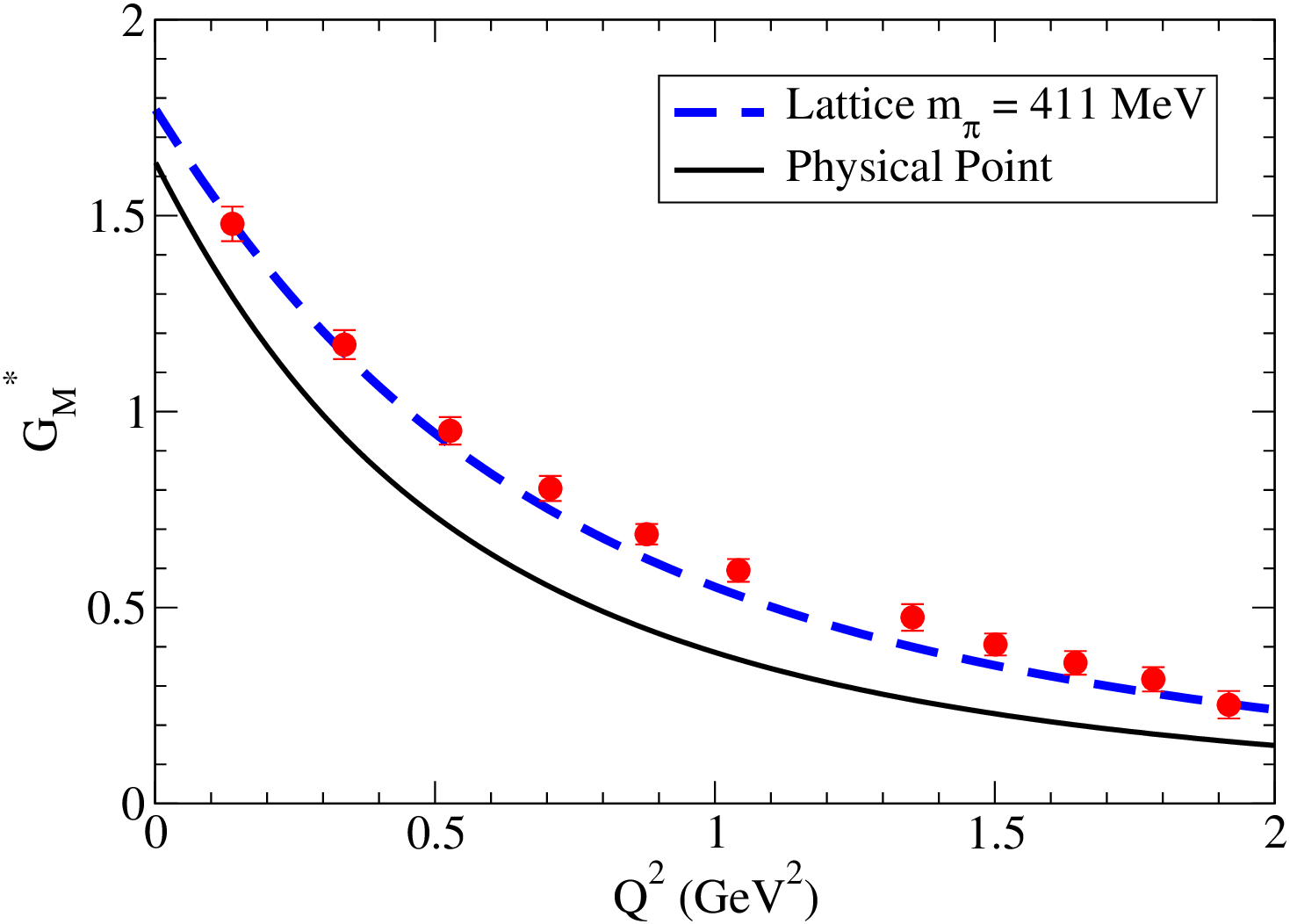}}}
\caption{Magnetic dipole form factor $G_M^\ast$ 
in quenched QCD \cite{Alexandrou08}.
The valence quark contribution at the 
physical point (solid line) is included 
as reference.}
\label{figGMfit}
\end{figure*}

\begin{table}[t]
\begin{center}
\begin{tabular}{c c c c c}
\hline
$m_\pi$ (GeV)   &  $\chi^2({G_M^\ast})$ & 
$\chi^2({G_E^\ast})$ & $\chi^2({G_C^\ast})$ & $\chi^2$ \\ 
\hline
0.563      &  0.569   & 0.483 &  0.853  & 0.618 \\   
0.490      &  0.544   & 0.290 &  0.668  & 0.513 \\
0.411      &  1.956   & 0.548 &  1.163  & 1.406 \\
\hline
Total      &          &       &         & 0.842 \\
\hline
\end{tabular}
\end{center}
\caption{Quality of the quenched fit in $\chi^2$ 
(partial and total) for the three 
form factors at the respective  pion mass. 
Quenched lattice QCD data from Ref.~\cite{Alexandrou08}.}
\label{tabCHI2}
\end{table}


\begin{table}[t]
\begin{center}
\begin{tabular}{c c c c}
\hline
$\alpha_2$,  &
  $\alpha_3$, $\alpha_4$ &   $a$, $b$ \\
\hline 
      0.3421 &  0.3507 & 0.0856  \\   
             &  0.3377 & 0.0857  \\  
\hline
\end{tabular}
\end{center}
\caption{
D-state parameters of the $\Delta$ wavefunction
as result of the fit to the quenched lattice data
\cite{Alexandrou08}.
The coefficient $\lambda_{D1}=1.0319$, in Eq.~(\ref{eqPsiD1})
is determined by the values of 
$\alpha_3$ and $\alpha_4$.}
\label{tabParam}
\end{table}

\begin{figure}[t]
\vspace{0.2cm}
\centerline{\mbox{
\includegraphics[width=2.4in]{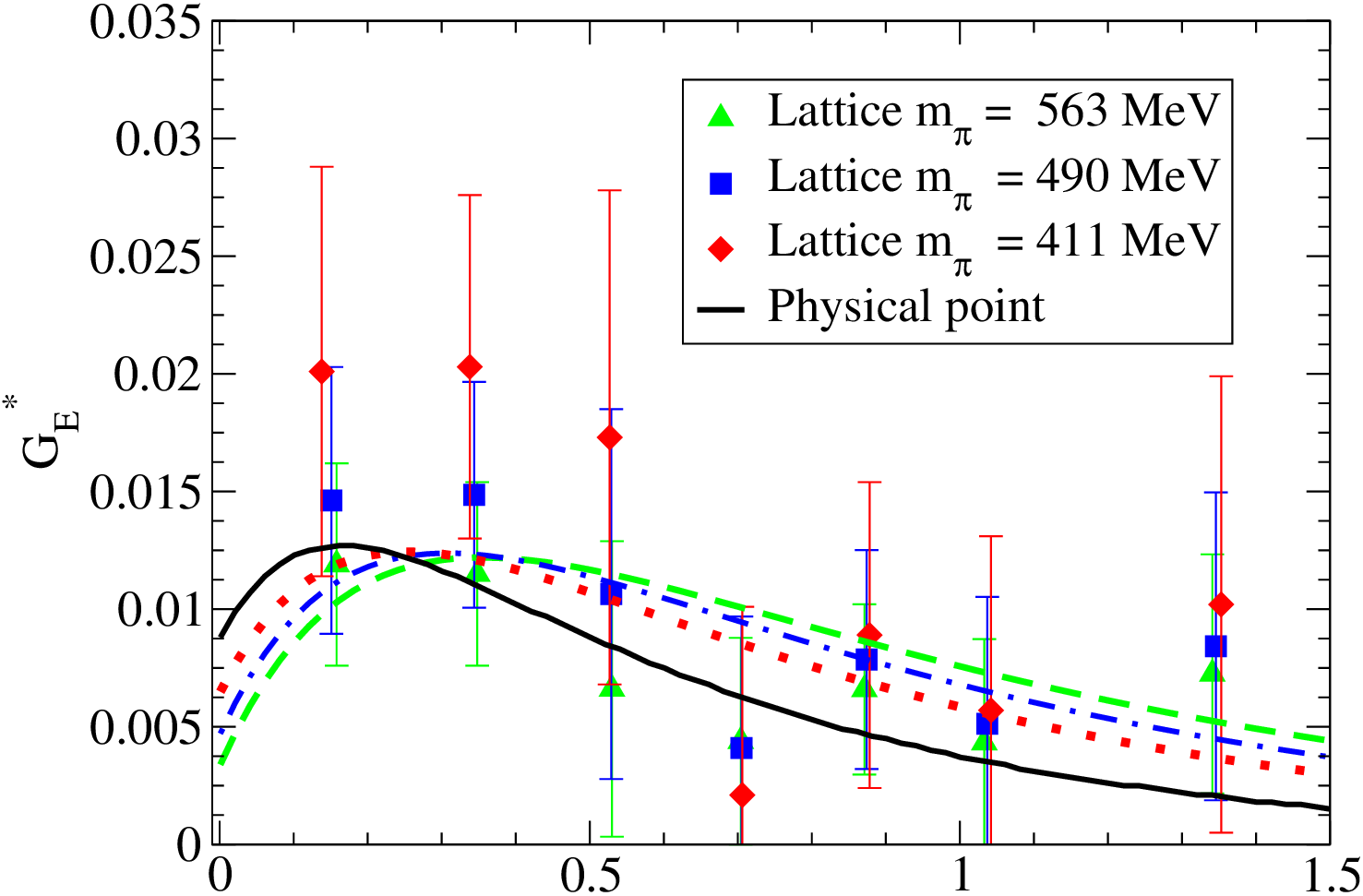}}}
\vspace{0.8cm}
\centerline{\mbox{
\includegraphics[width=2.4in]{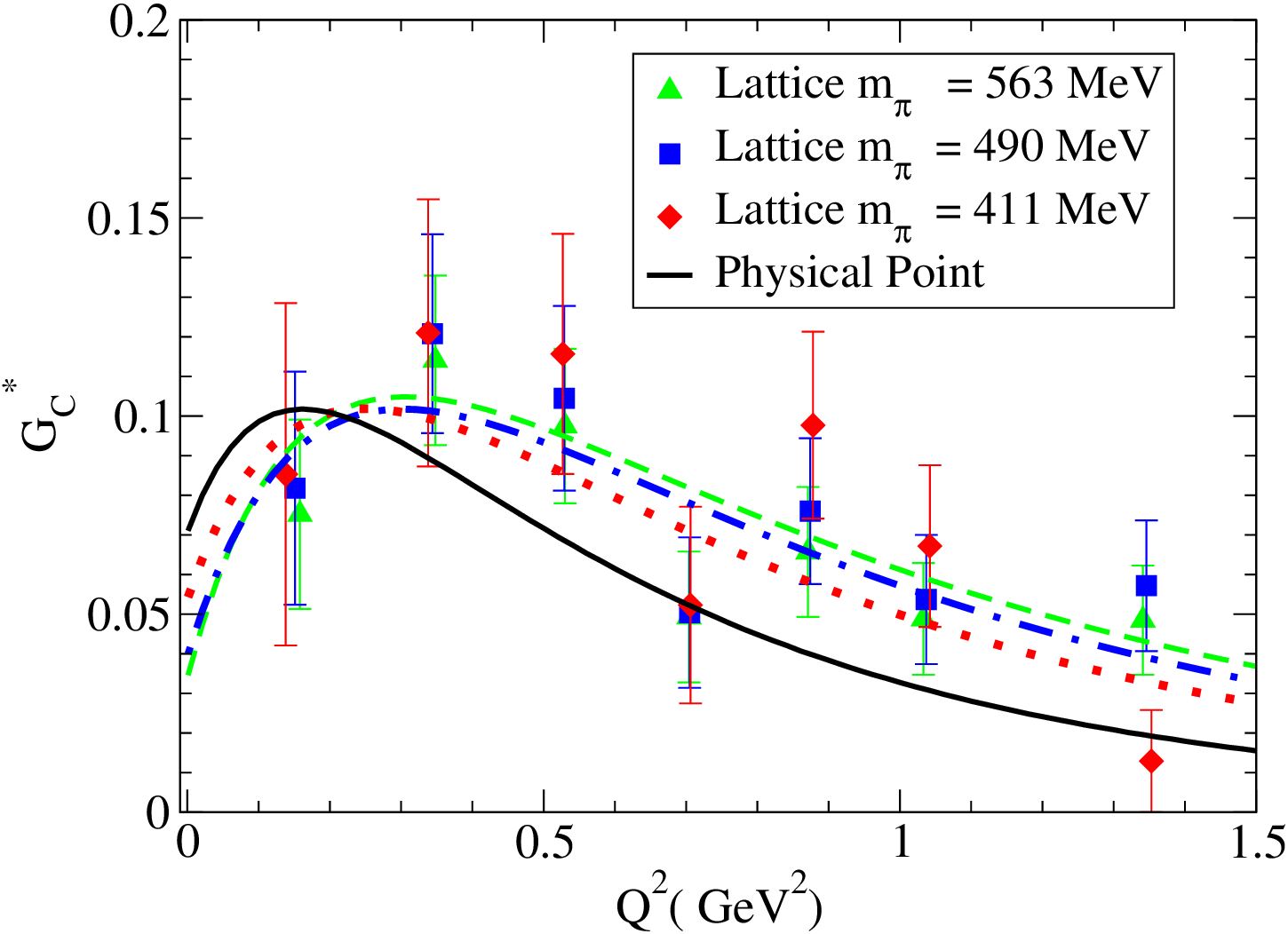}}}
\caption{Best fit of the quadrupole form factors in quenched 
lattice QCD \cite{Alexandrou08}.
The lines have the same meaning of the Fig.~\ref{figModelII}.}
\label{figGEGCfit}
\end{figure}

\section{Adjustment of the D-state 
parameters to the lattice data}
\label{lattice}

Given the results of the last section, 
we decided to change our strategy in the process
of comparing the quark model results with the lattice data.
Instead of trying to use the quadrupole 
data in the physical region, 
where valence and pion cloud contributions 
are both important, we fit first the lattice 
data and extrapolate to the
physical point.
With this procedure we avoid 
ambiguities related to the 
exact contribution of the pion cloud mechanisms and their 
entanglement with the D-wave effects.
The parameters which are varied in the fit are the admixture 
coefficients $a$ and $b$ and the three range 
parameters $\alpha_i$ ($i=2,3,4$), all associated to the D-state 
scalar wavefunctions.
We fit these five parameters to the lattice data.
%
The S-state parameter ($\alpha_1$)
was not fitted, and the value $\alpha_1= 0.3366$ from Ref.~\cite{NDeltaD} was used.

The partial and total results for the $\chi^2$ obtained 
are presented in the Table \ref{tabCHI2}.
The final results for the observables are presented in the 
Figs.~\ref{figGMfit} and \ref{figGEGCfit}.
The D-wave model parameters associated with the fit 
are presented in Table \ref{tabParam}.

The obtained range parameters are larger than the ones
from the fit to the physical data \cite{NDeltaD}, 
suggesting that the D-states are less 
peripheral (i.e., have a shorter range in configuration space) 
than inferred in Ref.~\cite{NDeltaD},
where we used an indirect estimate of the 
valence quark contribution at the physical pion mass point,
based on specific assumptions about the pion cloud contribution.
Another interesting point is that
the D-state range parameters $\alpha_i =0.338-0.351$, with $i=2,..,4$,
do not spread over a large region, 
suggesting one single value,  $\alpha_i\simeq 0.344$, as 
the signature range of the D-state regime, 
slightly larger than that of the  S-state range $\alpha_1\simeq 0.337$.
We note that, in this case,  it is the additional power 
in the D-state scalar wavefunctions
(see Eqs.~(\ref{eqPsiS})-(\ref{eqPsiD1})) 
that implies a more peripheral 
character for those states, 
when compared with the S states. Indeed,
as for low $k$, $\chi = \sfrac{k^2}{m_s^2}$ \cite{Gross06}, 
a higher power in momentum space corresponds to 
a more peripheral effect 
in configuration space, as it is to be
expected from a D-wave contribution.
The best fit corresponds to an admixture 
of 0.72\% for both  D3 and D1 states in the $\Delta$ wavefunction.

Compared to the results obtained by using the physical data alone,
the significant change occurs for the percentage of the D1 component,
that drops from 4.36\% to 0.72\%.
As for the state D3 the differences are minor.
In Ref.~\cite{NDeltaD} the percentage was 0.88\%.
As $a \simeq b$, we conclude that the quenched lattice data
is consistent with an equal admixture for both D-states.
The initial number of  effective    
parameters needed for a good fit 
can be reduced from five to four.

In Fig.~\ref{figGMfit} the 
quenched $G_M^\ast$ data is very well described 
by our model [which 
is also stated in the Table \ref{tabCHI2} in 
the column $\chi^2({G_M^\ast})$].
The exception is the case for the lightest pion mass
$m_\pi= 411$ MeV, where pion cloud effects may start 
to be important \cite{Detmold} but are absent in the valence
quark model.

The model describes fairly well the lattice 
data for the quadrupole form factors. 
The quality decreases for 
the lightest pion mass, which is due to the omission of explicit
pion cloud effects
in our approach.
It is encouraging that the $\chi^2$ values  
are lower than the ones found
for the fit in the physical region \cite{NDeltaD}, 
indicating that the procedure used here
is more natural. 
Still, it should be said that the 
$\chi^2$'s obtained  also possibly reflect
the still poorer quadrupole lattice statistics 
and the narrower range of the lattice data, 
when compared to the experimental data or even to
the lattice data for $G_M^\ast$.

A study of the dependence of the 
$\gamma N \to \Delta$ form factors 
on the pion mass and $Q^2$  
was also considered in Refs.~\cite{Pascalutsa06,Gail06}.
The lattice QCD data for $G_M^\ast$ 
in Refs.~\cite{Alexandrou,Pascalutsa06},
manifests a significant difference from the 
more recent analysis of Ref.~\cite{Alexandrou08}.
For similar pion mass the results 
of the Ref.~\cite{Alexandrou} 
are larger than the ones presented in Ref.~\cite{Alexandrou08}.

\section{Quadrupole form factors at the physical point}
\label{physical}

\begin{figure*}[t]
\centerline{\mbox{
\includegraphics[width=2.4in]{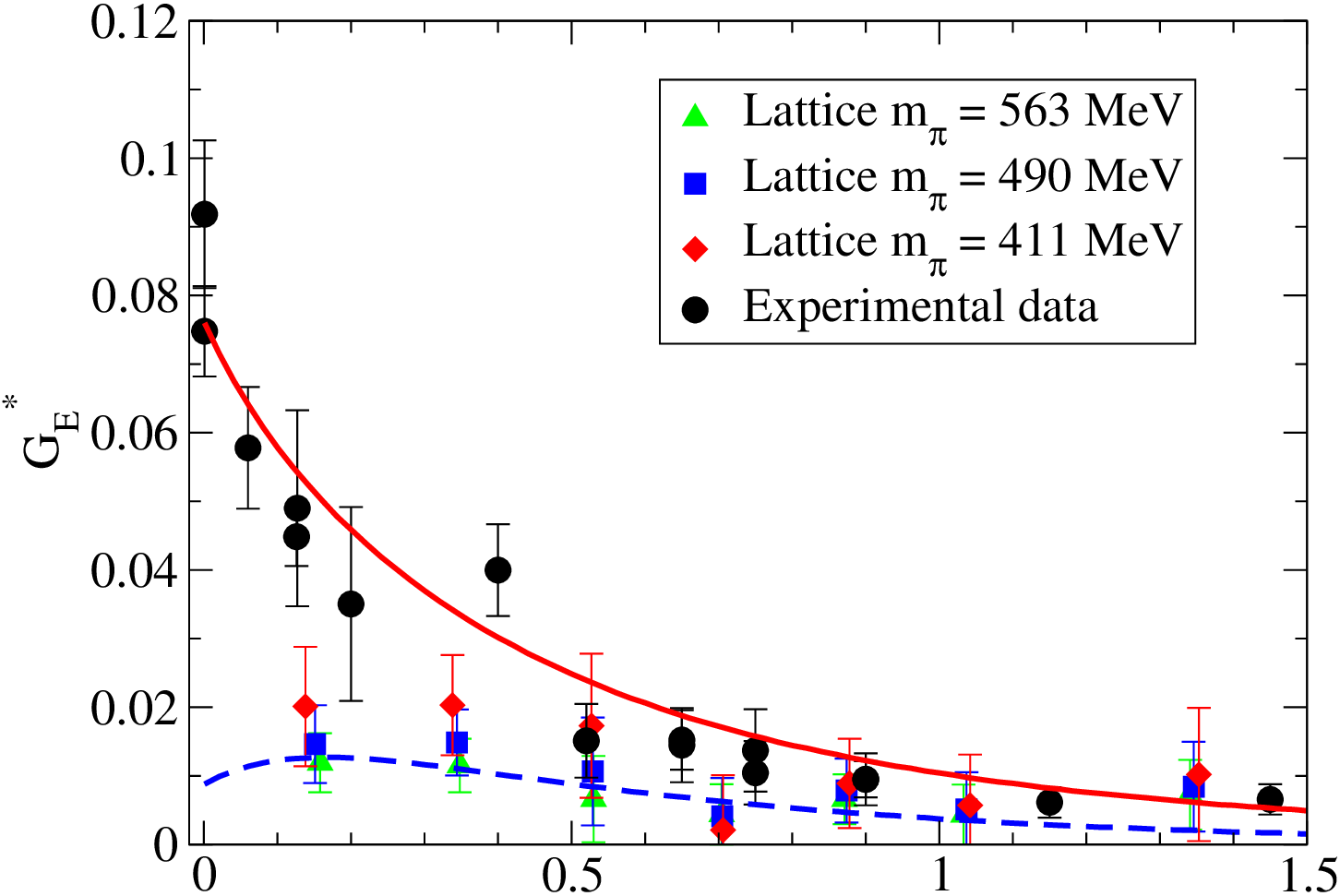}
\hspace{.3cm}
\includegraphics[width=2.4in]{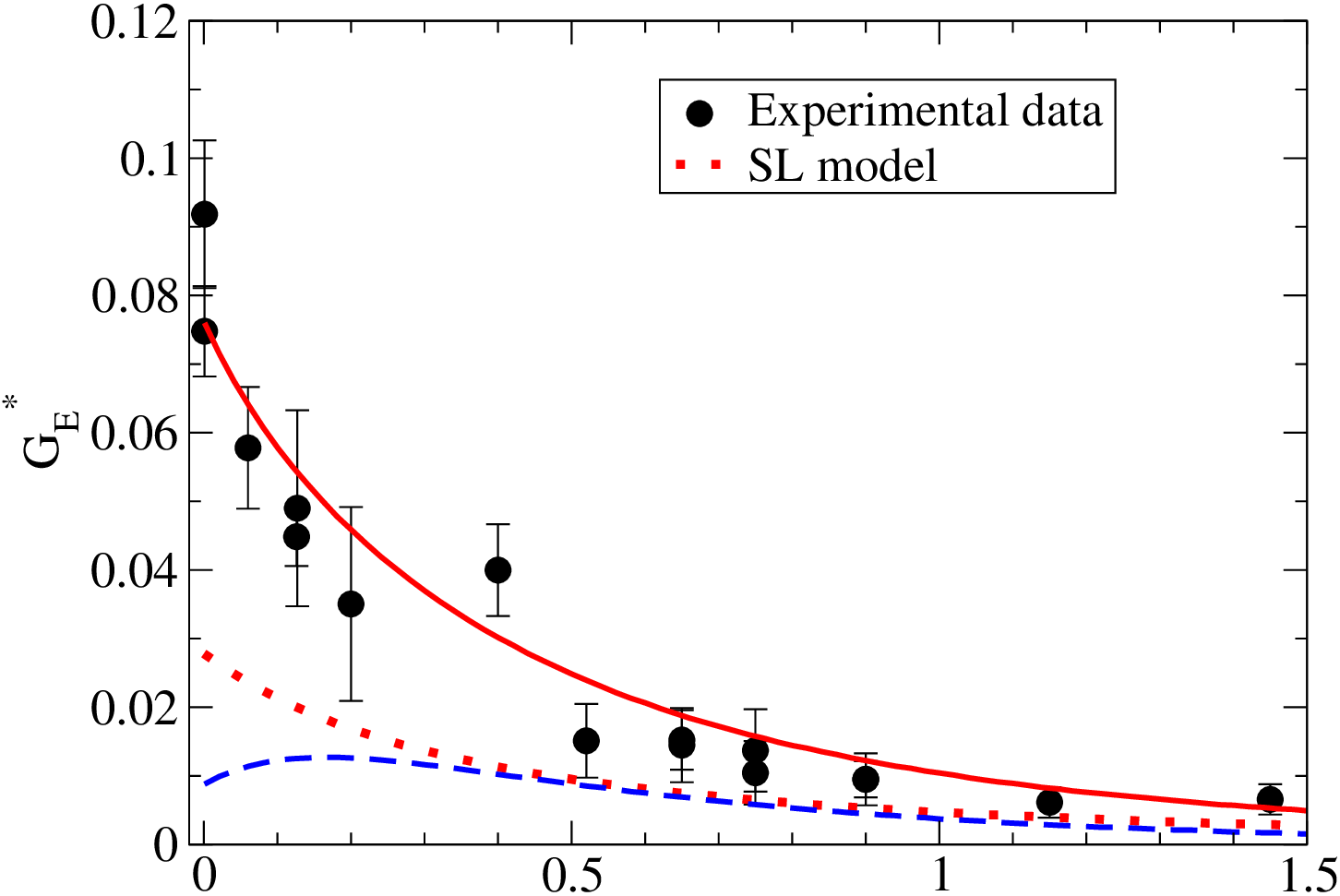}}}
\vspace{0.7cm}
\centerline{\mbox{
\includegraphics[width=2.4in]{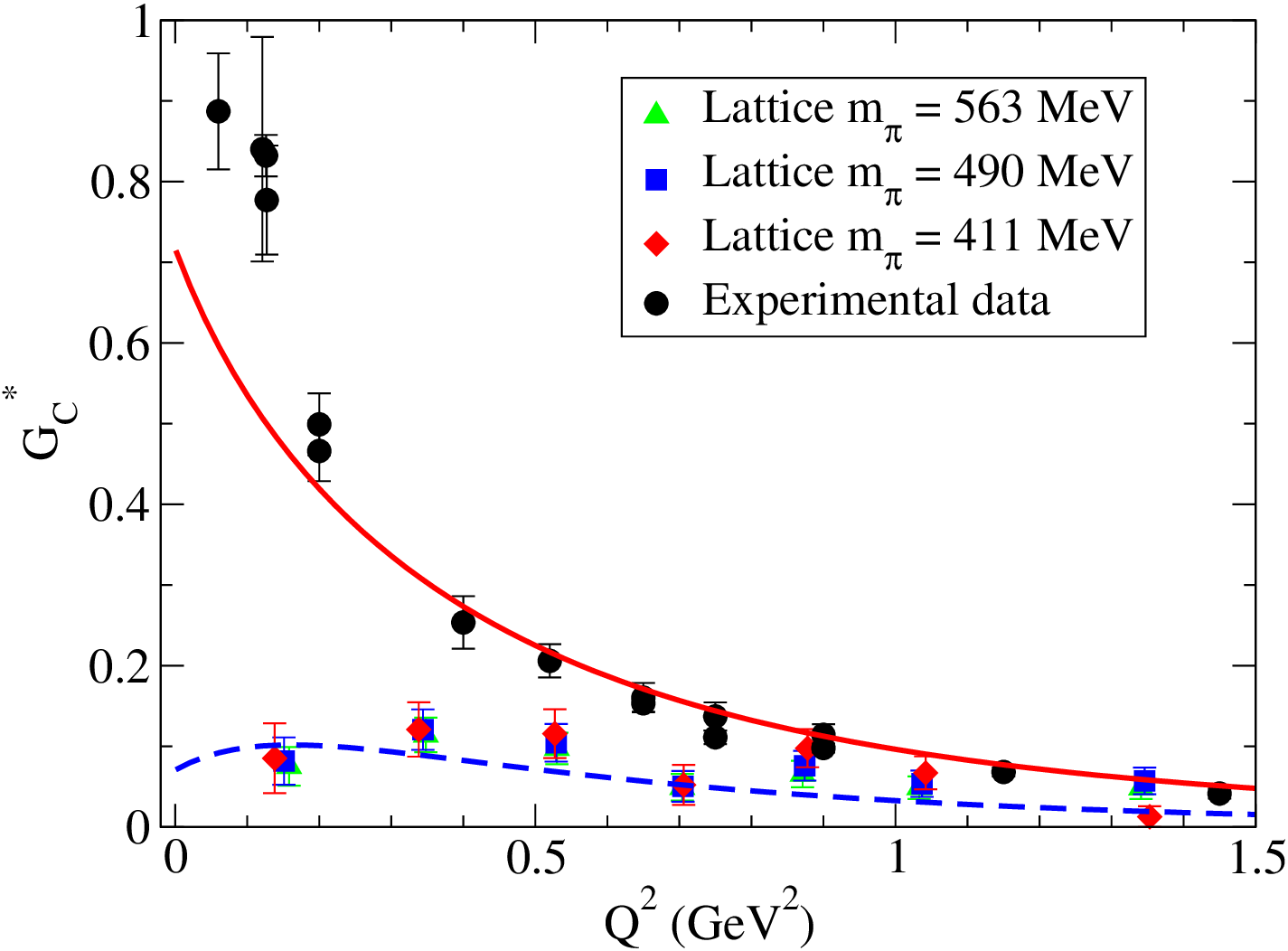}
\hspace{.3cm}
\includegraphics[width=2.4in]{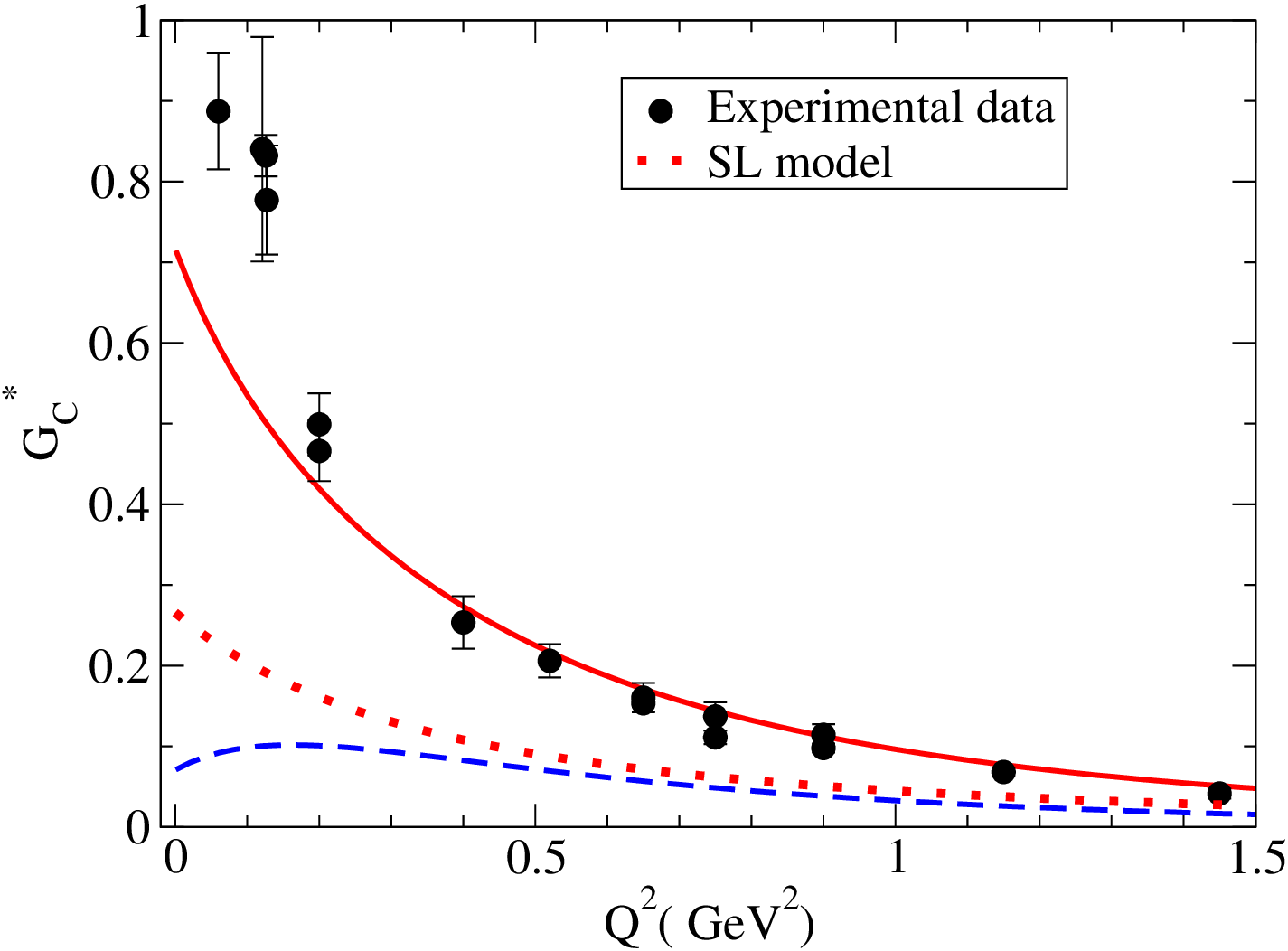}}}
\caption{Extension to the physical 
region using the quenched parametrization.
The dashed line represents the valence 
contribution.
The solid line represents the combination
of valence and pion cloud effects.
Physical data from Jlab \cite{CLAS,CLAS06}, 
MAMI \cite{MAMI}, LEGS \cite{LEGS} and
MIT-Bates \cite{Bates}.
Quenched lattice QCD data from \cite{Alexandrou08}.
Sato and Lee parametrization from \cite{Diaz07a}.}
\label{figGtotal}
\end{figure*}

After the parametrization of the D-states
was obtained from the lattice data, we can now apply it
to the physical region. This requires that we use the experimental data, in the form
of the two ratios
\be
R_{EM} = - \frac{G_E^\ast (Q^2)}{G_M^\ast(Q^2)},
\hspace{.3cm}
R_{EM} = - \frac{|{\bf q}|}{2 m_\Delta}
\frac{G_C^\ast (Q^2)}{G_M^\ast(Q^2)},
\ee
where $|{\bf q}|$ is the photon momentum in 
the $\Delta$ rest frame, 
and we use the empirical 
parametrization of Ref.~\cite{Gail06}:
\be
G_M^\ast(Q^2) = 3G_D \exp(-0.21 Q^2)
\sqrt{1+ 
\frac{Q^2}{(m_N+ m_\Delta)^2}}.
\ee
In the last expression $G_D=\left(1+Q^2/0.71\right)^{-2}$ 
represents the dipole form factor.
The quality of the parametrization was 
studied in the Ref.~\cite{NDeltaD}.

Since D-states extracted from the lattice data 
and applied to the physical region include only 
the contribution of the valence quarks,
they necessarily underestimate the experimental data. 
To fill the gap between the valence contribution 
and the experimental data we have to 
consider contributions from the pion cloud.
In particular we consider the pion cloud parametrization 
used in Ref.~\cite{NDeltaD},
where the pion cloud contributions 
to $G_E^\ast$ and $G_C^\ast$ 
were determined using  large-$N_c$ relations 
\cite{Pascalutsa07a,Buchmann07a}
between those form factors and 
$G_{En}$ (the neutron electric form factor):
\ba
& &
G_E^\pi (Q^2)=\left(\frac{m_N}{m_\Delta}\right)^{3/2} 
\frac{m_\Delta^2-m_N^2}{2 \sqrt{2}} \frac{G_{En}(Q^2)}{Q^2} \\
& &
G_C^\pi (Q^2)=\sqrt{\frac{2m_N}{m_\Delta}} m_N  m_\Delta 
\frac{G_{En}(Q^2)}{Q^2}. 
\ea
To evaluate $G_{En}$, we took 
model II of \cite{Nucleon} for the nucleon.
The results are presented in Fig.~\ref{figGtotal}.
In that figure we compare the final results 
for $G_E^\ast$ and $G_C^\ast$ and include
the lattice data to show the magnitude of the valence 
contributions.
The valence contributions are also 
compared with the parametrization 
of the valence contribution from the Sato and Lee model 
\cite{Diaz07a}. 
Note that the Sato and Lee parametrization 
gives a contribution similar to our model 
for $Q^2 > 0.5$ GeV$^2$, for both $G_E^\ast$ and $G_C^\ast$. 
It lies above our results, overpredicting the 
lattice data, for lower $Q^2$.

In conclusion, by fixing the D-state 
components by the lattice data
and considering a pion cloud parametrization,
derived from the large-$N_c$ limit at the physical point, 
we obtained a fairly good description of the quadrupole lattice data,
in the range $Q^2 < 1.5$ GeV$^2$. The exception is the region $Q^2 < 0.2$ GeV$^2$ 
where a small D1-mixture,
when compared with Ref.~\cite{NDeltaD}, 
underpredicts the $G_C^\ast$ data.
Note however that there is some discrepancy 
between different experimental data in that region \cite{NDeltaD}. 
The planned data from the CLAS collaboration for that range 
would be important to clarify 
the low $Q^2$ behavior of $G_C^\ast$
\cite{Diaz07a,Stave08}.

\begin{figure}[t]
\vspace{1.2cm}
\centerline{
\mbox{
\includegraphics[width=3.0in]{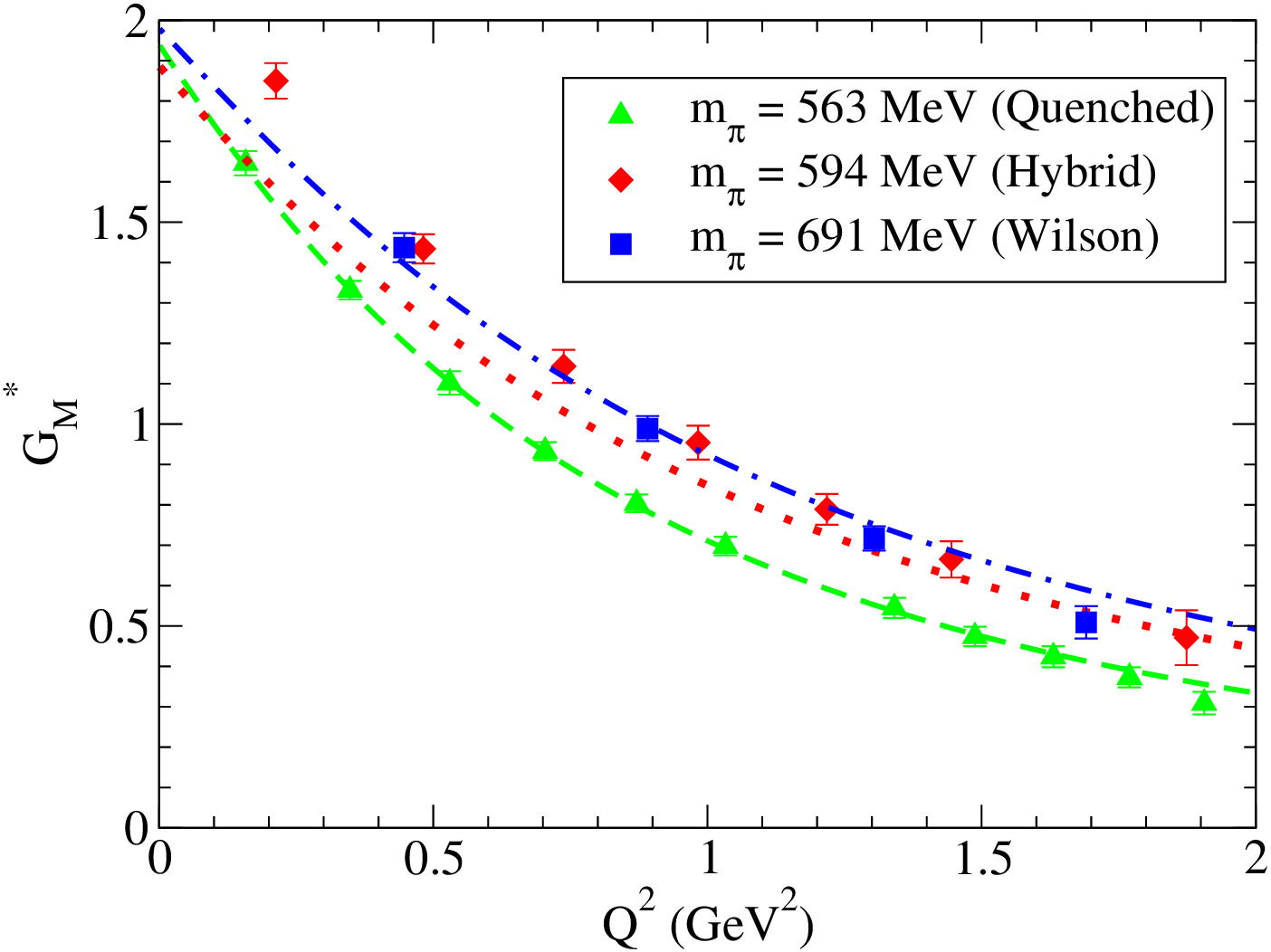} }}
\caption{Dependence of $G_M^\ast$ with $m_\pi$,
for quenched ($m_\pi = 563 $ MeV), 
Wilson ($m_\pi = 691$ MeV)  and Hybrid ($m_\pi = 594$ MeV).
The lines correspond to the result of the VMD model 
with $m_N$, $m_\rho$ and $m_\Delta$ 
associated to the quenched data for $m_\pi = 563$ MeV
(dashed), Wilson action: $m_\pi=594$ MeV (doted)
and Hybrid action: $m_\pi=691$ MeV (dashed-dotted).  }
\label{figGM600}
\end{figure}

A complete lattice QCD dataset is available 
in Ref.~\cite{Alexandrou08}. 
It  presents lattice data 
in the quenched approximation, 
and the unquenched data based on Wilson and also on a
hybrid action.
In this work we restrict our application to the quenched data.
There are three main reasons for this restriction:
\begin{itemize}
\item
There is a significant discrepancy 
between the quenched and unquenched data, 
particularly for the results of $G_M^\ast$ 
with heavier masses. In this regime we would expect small
pion cloud effects, implying negligible 
differences between quenched and unquenched results.
In Fig.~\ref{figGM600} we compare the lattice data 
corresponding to $m_\pi= 563$ MeV (quenched) and  
$m_\pi= 594$ MeV (hybrid action). 
There is a significant difference 
between those two data sets, with 
the Wilson data associated with  $m_\pi=$ 691 MeV
being more consistent with the hybrid action ($m_\pi= 563$ MeV).
\item
There are differences between 
the two unquenched results, in particular 
for the value of $m_\rho$.
The extension of our
model depends on the (quenched) $\rho$  mass. 
It is not clear whether the extension of 
our model is justified 
for the unquenched calculations,
where the nucleon, $\Delta$, and $\rho$ 
masses would differ from the quenched masses.
We would expect only minor differences 
for heavier pion masses (say $m_\pi > 480$ MeV). 
However, the 
significant difference between 
the $\rho$ mass for the Wilson action data with $m_\pi=509$ MeV 
($m_\rho=887$ MeV) and the hybrid action with $m_\pi=490$ MeV ($m_\rho= 949$ MeV)
is difficult to explain. 
\item
Finally, there is only
a limited number of unquenched quadrupole data points 
for large pion masses. Since 
the unquenched lattice data for light 
pion masses as 353 MeV (hybrid action) and 
384 MeV (Wilson action) are expected to be 
contaminated with pion cloud 
effects, which cannot be simulated by 
our valence quark model, 
those points would have to be excluded. We would then be left 
with 6 or 8-9 quadrupole points, 
respectively, for the Wilson and hybrid action 
(to be compared with 21 from quenched data),
and with such a small number of constraints 
the fit would naturally become meaningless.
\end{itemize}

Apart from the disagreement 
observed in the description of the magnetic 
dipole form factor, 
the Wilson and hybrid action lattice data
suggest a weaker falloff of the 
electric quadrupole form factor $G_E^\ast$ 
when compared to the quenched prediction.
This result is also observed for the 
physical point extrapolation.

Once the differences between 
the two unquenched results are understood,
and the disagreement
between quenched and unquenched results
for $m_\pi \sim 600$ MeV is clarified,
it would be interesting to use 
also unquenched data to extract the
contribution of the D-states, 
using the procedure suggested here.
The increasing of statistics in both 
quenched and unquenched lattice data 
could also help to constrain the effects 
of the $\Delta$ D-states in the $\gamma N \to \Delta$ 
transition.

\section{Conclusions}
\label{conclusions}

In this work we study the valence 
quark contributions to the $\gamma N \to \Delta$ transition
in the lattice QCD regime, in
the framework of the
covariant spectator formalism. 
The nucleon and the $\Delta$ wavefunctions are not derived 
from a wave equation but are parametrized in terms of the nucleon and $\Delta$
symmetry structure for spin, isospin and angular momentum. 
By construction, the formalism
includes only  valence quark degrees of freedom, whereas
meson cloud mechanisms are not 
taken into account.

As discussed extensively in the literature
(see Refs.~\cite{NDeltaD,Pascalutsa06}), valence contributions 
do not dominate the quadrupole form factors at the physical point,
where the  pion cloud effects dominate instead, 
while the opposite happens in the 
lattice QCD regime.
In an attempt to explain the 
significant difference between the 
experimental data and the emerging 
simulations of lattice QCD with decreasing 
pion mass, but still small chiral effects 
associated with the light pions, 
we started by comparing our quark model
directly to the lattice data.
In contrast to the dominant contribution 
controlled by the nucleon and $\Delta$ S-states,
the $\Delta$ D-states, being a second order correction, 
are more sensitive to the lattice data.
This sensitivity provides a clean evaluation
of the valence quark contribution, since in the physical region the 
valence quark contribution is masked by the
overwhelmingly larger contribution of the pion cloud, and consequently
the D-state parametrization cannot 
be accurately constrained.

Accordingly, we found that by fixing the D-states by the physical data
first does not lead to a good description of the lattice QCD data.
We also verified that, inversely, when the D-states 
are first fixed in the lattice QCD regime,
then a good description of the physical data is possible.
In fact, adding the valence quark contribution
extrapolated from quenched lattice QCD to the physical 
mass regime, with an estimate of the pion cloud 
based on the large-$N_c$ limit \cite{NDeltaD,Pascalutsa07a},
the experimental data for $\gamma N \to \Delta$ quadrupole 
form factor is well described. An even more accurate 
description of the experimental data 
can in principle be obtained by considering 
more and more precise lattice QCD data,
and a more sophisticated estimate
of the pion cloud.

The fit to the lattice QCD data varied five parameters
associated with the valence D-state states.
The result of the fit suggests 
an identical admixture of the D1 and D3 states.

We conclude that lattice QCD data are important  
to constrain valence quark models.
Since for lattice calculations with $m_\pi > 400$ MeV  
valence quark effects dominate over the pion cloud, 
lattice data can be used to study and separate those effects.

\vspace{0.3cm}
\noindent
{\bf Acknowledgments}

G.~R.~wants to thank Jozef Dudek, Kostas Orginos 
and Franz Gross for the helpful discussions. 
The authors thank Constantia Alexandrou 
for sharing details of the lattice data 
presented in Ref.~\cite{Alexandrou08} 
and Alfred Stadler for the review of the final text.
This work was partially support by Jefferson Science Associates, 
LLC under U.S. DOE Contract No. DE-AC05-06OR23177.
G.~R.\ was supported by the portuguese Funda\c{c}\~ao para 
a Ci\^encia e Tecnologia (FCT) under the grant  
SFRH/BPD/26886/2006. 
This work has been supported in part by the European Union
(HadronPhysics2 project ``Study of strongly interacting matter'').

\end{document}